\documentclass[12pt]{iopart}

\usepackage{iopams}

\expandafter\let\csname equation*\endcsname\relax
\expandafter\let\csname endequation*\endcsname\relax
\usepackage{amsmath}  
\usepackage{graphicx}
\begin{document}

\title{Fermions tunneling of Kerr-Newman-de Sitter black hole in Lorentz violation theory}

\author{ N. Media$^*$,  Y.\, Onika \,Laxmi$^*$, T. Ibungochouba Singh$^{*+}$}

\address{$^*$Mathematics Department, Manipur University, Canchipur}
\ead{$^+$ibungochouba@rediffmail.com}
\vspace{10pt}
\begin{indented}
\item[]
\end{indented}

\begin{abstract}
In this paper, the tunneling of fermions near the event horizon of Kerr-Newman-de Sitter black hole is investigated in frame dragging coordinate systems, Eddington coordinate system and Painleve coordinate system  by using Dirac equation with Lorentz violation theory, Feynman prescription and WKB approximation. The Hawking temperature, heat capacity and change in Bekenstein-Hawking entropy are modified due to presence of Lorentz violation theory. The modified Hawking temperatures, heat capacities and change in Bekenstein-Hawking entropies near the event horizon of Kerr-Newman-de Sitter black hole would increase or decrease depending upon the choices of ether like vectors $u^{\alpha}$. In the absence of Lorentz violation theory, the original Hawking temperature, entropy and heat capacity are recovered.
\end{abstract}

%
\noindent{\it Keywords}: Kerr-Newman-de Sitter black hole, Dirac equation, Lorentz symmetry violation, Ether like vectors $u^{\alpha}$\\
PACS number: 04.62.+v, 04.70.Dy
%
%
%

\section{Introduction}
Hawking proposed the emission of black hole using quantum field theory in curved space time [1, 2]. The relation between the radiation of black hole and black body radiation have been completely established as $T=k/2\pi$, where $k$ is the surface gravity near the event horizon of black hole [3]. Refs. [4, 5] showed that the entropy of black hole is proportional to horizon area. Since then, many scientists have proposed different methods to find out the Hawking radiation for the stationary and non stationary black holes. Refs. [6, 7 ] proposed a new method known as tortoise coordinate transformation in which gravitational field is not dependent on time. In this method, the radial part of Klein-Gordon equation, Dirac equation and Maxwell electromagnetic field equations of stationary and nonstationary black holes are obtained by using tortoise coordinate transformation and null surface equation. Thus the incoming wave and the outgoing wave are obtained and the corresponding thermal radiation spectrum could be investigated. Applying their method; many interesting results of black hole radiation are obtained in [8-15 ]. Parikh and Wilczek [16-18] discussed the derivation of Hawking radiation as a tunneling process across the event horizon of the black hole. They obtain the potential barrier is created by the outgoing particle which is related to the Boltzmann factor for emission at the Hawking temperature. Their main aim is to construct a well-behaved coordinate system near the event horizon to find the emission rate. Refs. [19-21] have extended the Parikh-Wilczek's method to study the tunneling of charged massive particle at the event horizon of black hole. Kerner and Mann [22] proposed the tunneling of fermions near the event horizon of black hole by using Dirac equation, Pauli Sigma matrices and Feynman pescription. Applying the tunneling method to Rindler spacetime, they derive the Unruh temperature. Ref. [23] as an extension of Srinivasan and Padmanabhan [24] discussed the Hawking radiation as a tunneling near the event horizon of black hole by using Hamilton-Jacobi equation, Feynman prescription and WKB approximation. They showed that the naive coordinate gave the incorrect Hawking temperature whereas well-behave coordinate gave the original Hawking temperature.  Banerjee and Majhi [25-28] proposed the Hawking as tunneling across the event horizon of black hole by using Hamilton-Jacobi equation beyond the semiclassical approximation. They derived the entropy correction of black hole containing logarithmic term by using first law of black hole thermodynamics. Following their method, entropy corrections in different black holes have been obtained in [29-31]. The tunneling of vector boson particle near the event horizon of black hole is investigated by using Hamilton-Jacobi ansatz to Proca equation, WKB approximation and Feynman prescription [32, 33]. Applying above method, the Hawking temperatures in different black holes have been derived in [34-37]. The Lorentz symmetry which is the cornerstone of general theory of relativity may break at high energy. Refs. [38-43] proposed the various gravity models based on Lorentz symmetry violation. The problem of irrenormalization of gravity theory can be solved by using Lorentz symmetry violation. Mukohyama [44] suggested that the theory of dark matter might be the effect of theoretical model of Lorentz violation theory. Refs. [45-47] discussed the Lorentz symmetry violation in the fields of string theory, electrodynamics and non-abel theory. Refs. [48, 49] discussed the Lorentz symmetry violation by using Dirac equation in flat space time with ether like field terms. They investigated the quantum correction of ether-like field terms. It is noted that the presence of ether like field tends to the Lorentz symmetry violation of the space time. The properties which are not consistent with Lorentz symmetry violation theory may appear at high energy. Taking Lorentz symmetry violation into account, the dynamics of fermions in curved space time and the modification of quantum tunneling radiation caused by ether like field may be worthy of study. Ref. [50] studied the quantum tunneling radiation of spherically symmetry black holes by using Dirac particle with ether like vectors. Applying Lorentz violation theory, the modified Hawking radiations and entropies of different black holes have been discussed in [51-56].

 The aim of this paper is to investigate the modified Hawking temperatures, heat capacities and change in Bekenstein-Hawking entropies near the event horizon of Kerr-Newman-de Sitter black hole in frame dragging coordinates, well-behaved Eddington coordinate and Painleve coordinate with Lorentz violation theory in curved space time. Choose appropriate ether like vectors $u^{\alpha}$ from the line element of Kerr-Newman-de Sitter black hole, we show that the Hawking temperatures, heat capacities and change in Bekenstein-Hawking entropies near the event horizon of KNdS black hole will be affected by the choices of ether like vectors $u^{\alpha}$.
 
 The organization of this paper is as follows. In section 2, the original Hawking temperature and the heat capacity near the event horizon of Kerr-Newman-de Sitter (KNdS) black hole are derived. We discuss the modified Hawking temperatures, heat capacities and change in Bekenstein-Hawking entropies of KNdS black hole in 3-dimensional and 4-dimensional frame dragging coordinates using Dirac equation with Lorentz violation theory in curved space time in sections 3 and 4 respectively. Using Lorentz violation theory in curved space time, the modified Hawking temperature, heat capacity and change in Bekenstein-Hawking entropy of KNdS black hole are investigated in well-behave Eddington coordinate in section 5. In section 6, we discuss the modified Hawking temperature, heat capacity and change in Bekenstein-Hawking entropy of KNdS black hole in well-behaved Painleve coordinate using Dirac equation with Lorentz violation theory in curved space time. Some discussions and conclusions are given in section 7. 
 
\section{Kerr-Newman-de Sitter black hole}

The line element of the Kerr-Newman-de Sitter black hole in Boyer-Lindquist coordinate $(t,r,\theta,\phi)$ [57] is given by 
\begin{align}
ds^2&=-\frac{\Delta_r-\Delta_\theta a^2\sin^2\theta}{\rho^2\Xi^2}dt^2+\frac{\rho^2}{\Delta_\theta}d\theta^2
-\frac{2a[\Delta_\theta(r^2+a^2)-\Delta_r]\sin^2\theta}{\rho^2\Xi^2}dtd\varphi \cr&+\frac{\rho^2}{\Delta_r}dr^2+\frac{\Delta_\theta(r^2+a^2)^2-\Delta_r a^2\sin^2\theta}{\rho^2\Xi^2}\sin^2\theta d\varphi^2,
\end{align}
where
\begin{eqnarray}
&&\rho^2=r^2+a^2\cos^2\theta,\,\,\,\,\, \Xi=1+\frac{1}{3}\Lambda a^2,\cr
&&\Delta_\theta=1+\frac{1}{3}\Lambda a^2 \cos^2\theta,\cr
&&\Delta_r=(r^2+a^2)\Big(1-\frac{1}{3}\Lambda r^2 \Big)-2Mr+q^2.
\end{eqnarray}
Here $M$, $a$ and $\Lambda$ denote the mass of the black hole,  rotational parameter and cosmological constant respectively.
Eq. (1) describes an interesting charge rotating Kerr-Newman-de  Sitter (KNdS) or Kerr-Newman-anti-de Sitter (KNAdS) black hole for $\Lambda >0$ or $\Lambda<0$ respectively. The singularity of KNdS black hole could be calculated from the null surface equation [12]
\begin{eqnarray}
g^{ab}\frac{\partial F}{\partial x^a}\frac{\partial F}{\partial x^b}=0,\end{eqnarray}
 where $F=F(r)$. The horizon equation is 
\begin{eqnarray}
\Delta_r = r^2-\frac{1}{3}\Lambda r^4 +a^2-\frac{1}{3}\Lambda r^2 a^2 -2Mr+q^2=0.
\end{eqnarray}
If $ a^2+q^2 < M^2 \leq \frac{1}{\Lambda}$, Eq. (4) gives four apparent singularities of KNdS black hole. The four roots are denoted by $r_a, r_h , r_b$ and $r_c $ $(r_a> r_h> r_b> r_c)$. $r_a, r_h$ and $r_b$ represent the location of cosmological horizon, event horizon and Cauchy horizon respectively. If one comes to the ring singularity $r=0,\theta=\frac{\pi}{2}$, then on the other side of $r=0, r=r_c$ is taken as another cosmological horizon [58]. 
To study the modified Hawking radiation near the event horizon of KNdS black hole,  we factorise $\Delta_r$ as $\Delta_r=(r-r_h)\Delta^{'}(r_h)$, where $r_h$ is defined by [59]
\begin{eqnarray}
  r_h&=&\frac{1}{\alpha_1}\left(1+\frac{4\Lambda M^2}{3\beta ^2\alpha_1}+...\right)\left(M+\sqrt{M^2-(a^2+q^2)\alpha_1}\right), 
\end{eqnarray}
  where  $\beta=\sqrt{1-\frac{\Lambda}{3} a^2} $ and $\alpha_1=\sqrt{(1+\frac{\Lambda a^2}{3})^2+\frac{4\Lambda q^2}{3}}$.
The KNdS black hole has a frame dragging effects of the coordinate system. To apply frame dragging effect, let $d\phi=-\frac{g_{14}}{g_{44}}dt$ and $\Omega=-\frac{g_{14}}{g_{44}}$, then Eq. (1) can be written as
\begin{eqnarray}
 ds^2&=& \hat{g}_{00}dt^2+g_{11}dr^2+g_{22}d\theta^2,
\end{eqnarray}
where $\hat{g}_{00}=-\frac{\Delta_r\Delta_{\theta}\rho^{2}}{\Xi^{2}[\Delta_{\theta}(r^2+a^2)^2-\Delta_ra^2 \sin^2\theta]}$, $g_{11}=\frac{\rho^2}{\Delta_r}$, $g_{22}=\frac{\rho^2}{\Delta_{\theta}}$.  The angular velocity at the event horizon of KNdS black hole is given as
\begin{eqnarray}
 \Omega &=&\frac{a}{r_h^2+a^2}.
 \end{eqnarray}
 
Eq. (6) satisfies the Landau's condition of coordinate clock synchronization. In such case the event horizon of KNdS black hole and infinite redshift surface are concordant. This indicates that the geometrical optics limit can be applied. The tunneling probability and the imaginary part of the action can be calculated by using WKB approximation [60].
	
The original Hawking temperature of KNdS black hole at the event horizon is defined by [61]
\begin{eqnarray}
T_H=\frac{1}{2\pi} \Big[\frac{r_h-M-\frac{2}{3}\Lambda r_h^3-\frac{1}{3}\Lambda r_ha^2}{\Xi (r_h^2+a^2)} \Big].
\end{eqnarray}
The original Bekenstein-Hawking entropy near the event horizon of KNdS black hole is calculated as
\begin{eqnarray}
S_{bh}=\frac{\pi (r_h^2+a^2)}{\Xi}.
\end{eqnarray}
To obtain the specific heat of KNdS black hole, the black hole mass can be derived from $\Delta(r_h)=0$ as
\begin{eqnarray}
M=\frac{r_h}{2}+\frac{a^2}{2r_h}+\frac{q^2}{2r_h}-\frac{\Lambda r_h^3}{6}-\frac{\Lambda a^2 r_h}{6}.
\end{eqnarray}	
The heat capacity $(C_h)$ of KNdS black hole near the event horizon is given [62] by
\begin{eqnarray}
C_h=\frac{\partial M}{\partial T_h}=\Big(\frac{\partial M}{\partial r_h }\Big) \Big(\frac{\partial r_h}{\partial T_h}\Big).
\end{eqnarray}
Using Eqs. (8) and (10) into Eq. (11), the heat capacity near the event horizon of KNdS black hole is obtained as
\begin{eqnarray}
C_H=\frac{2\pi \Xi (r_h^2+a^2)^2[3(r_h^2-a^2-q^2)-\Lambda r_h^2(3r_h^2+a^2)]}{3(a^4-r_h^4)+4a^2 r_h^2(3-2 \Lambda r_h^2)-\Lambda r_h^2(3r_h^4+a^4)+3q^2(3r_h^2+a^2)}.
\end{eqnarray}
\section{Lorentz violation theory in curved space time}
The particle action and the Dirac equation in flat space time with Lorentz violation theory are investigated by Nascimento et al. [48]. The Lorentz symmetry violation term is taken into account, the Dirac equation in a flat space time can be derived by using Hamilton principle. To extend the Dirac equation from flat space time to the KNdS black hole, the following two points are needed. First, one needs to generalise the ordinary derivatives of flat space time to the covariant derivative of curved  space time. Secondly, the commutation relation $\gamma^\mu$ from the flat space time should be extended to the commutation relation of curved space time. The Dirac equation of curved space time based on Lorentz symmetry violation theory with mass $m$ is given by [63]
 
 \begin{eqnarray}
 {\gamma^\mu D_\mu [1+\hbar^{2} \frac{a_k}{m^2}(\gamma^{\mu}D_{\mu})^2]+\frac{b}{\hbar}\gamma^5+c\hbar (u^\alpha D_\alpha)^2-\frac{m}{\hbar}}\psi=0,
 \end{eqnarray}
 where $a_k,b$ and $c$ denote the arbitrary small quantities and $0<\frac{a_k}{m}, \frac{b}{m},\frac{c}{m}<1$. $u^\alpha$ is the ether like vectors and other parameters satisfy the following conditions
 \begin{eqnarray}
 &&D_\mu =\partial_\mu +\frac{1}{2}i\Gamma^{\alpha \beta}_{\mu}\Sigma_{\alpha \beta}, \,\,\,\,\,\,\Gamma^{\alpha \beta}_{\mu}= g^{\beta \nu}\Gamma^{\alpha}_{\mu \nu},\cr
 &&\{{\gamma^\mu ,  \gamma^\nu}\} = 2g^{\mu\nu}I, \,\,\,\,\,\,\,\,\sum_{\alpha \beta}=\frac{1}{4}[\gamma^{\alpha},  \gamma^{\beta} ].
\end{eqnarray}
  From Eq. (6), the components of $\gamma^\mu $ can be constructed as
\begin{eqnarray}
\gamma^t &=&\frac{1}{\sqrt{\hat{g}_{00}}}
 \begin{bmatrix} 
	i&0\\
	0&-i\\
	\end{bmatrix},
	\nonumber\\
	\gamma^r &=&\frac{1}{\sqrt{g_{11}}}
 \begin{bmatrix} 
	0&\sigma^3\\
	\sigma^3&i\\
	\end{bmatrix},
	\nonumber\\
	\gamma^\theta &=&\frac{1}{\sqrt{g_{22}}}
 \begin{bmatrix} 
	0&\sigma^1\\
	\sigma^1&i\\
	\end{bmatrix},
	\quad
	\end{eqnarray}
	where $\sigma^i (i=1,3)$ denote the Pauli Sigma matrices. To investigate the tunneling of  spin-$\frac{1}{2}$ fermions, the wave function $\psi$ is taken as 
	\begin{eqnarray}
	\psi=\rm {S_0\,\,  exp}[\frac{i}{\hbar} S(t,r,\theta)],
	\end{eqnarray}
	where $S_0$ and $S$  are column matrix and Hamilton principal function.

Substituting Eqs. (6), (14), (15) and (16) in Eq. (13) and after some simplifications, the Hamilton principal function of spin-$\frac{1}{2}$ fermions in KNdS black hole is obtained as
\begin{eqnarray}
&&\left[ g^{tt}\left(\frac{\partial S}{\partial t}\right)^2+g^{rr}\left(\frac{\partial S}{\partial r}\right)^2+g^{\theta \theta}\left(\frac{\partial S}{\partial \theta}\right)^2\right](1+2a_k)+2cm u^t u^t\left(\frac{\partial S}{\partial t}\right)^2\cr &&+4cm u^t u^r \left(\frac{\partial S}{\partial t}\right)\left(\frac{\partial S}{\partial r}\right) +4cm u^t u^\theta \left(\frac{\partial S}{\partial t}\right)\left(\frac{\partial S}{\partial \theta}\right)+2cm u^r u^r \left(\frac{\partial S}{\partial r}\right)^2\cr &&+4cm u^r u^\theta \left(\frac{\partial S}{\partial r}\right)\left(\frac{\partial S}{\partial \theta}\right)+2cm u^\theta u^\theta \left(\frac{\partial S}{\partial r}\right)^2+ m^2=0.
\end{eqnarray}
To investigate the modified Hawking radiation from Eq. (17), we should choose the proper expression of $u^t,u^r$ and $u^\theta$. The  ether like vectors $u^\alpha$ are constant in flat space time but they are not constant in curved space time. We can claim $ u^{\alpha}u_{\alpha}={\rm constant}$. It is well known that $u^\alpha$ is related to the components of covariant of metric tensor.
The ether like vectors $u^\alpha$ from Eq. (6) can be constructed as follows
\begin{eqnarray}
   u^t&=&\frac{c_t}{\sqrt{-g_{tt}}} = \frac{c_t \Xi \sqrt{\Delta_{\theta}(r^2+a^2)^2-\Delta_r a^2 \sin^2\theta}}{\sqrt{\Delta_r\Delta_{\theta}\rho^{2}}} ,\cr
u^{r}&=&\frac{c_{r}}{\sqrt{g_{rr}}} =c_r\sqrt{\frac{\Delta_r}{\rho^2}},\cr
u^{\theta}&=&\frac{c_{\theta}}{\sqrt{g_{\theta\theta}}} =c_\theta \sqrt{\frac{\Delta_\theta}{\rho^2}}, 
 \end{eqnarray} where $c_t$, $c_r$ and $c_\theta$ are arbitrary constants.
   The three ether like vectors $u^t$, $u^r$ and $u^\theta$ satisfy the following condition \begin{eqnarray}
     u^{\alpha}u_{\alpha}=-c_t^2+c_r^2+c_\theta^2={\rm constant}.
\end{eqnarray}
 
The difficulty of solving Eq. (17) lies in the fact that it contains the variables $t,r$ and $\theta$. To solve Eq. (17), the action $S$ can be defined as
\begin{eqnarray}
 S=-\omega t+R(r,\theta) +j \phi+W,
\end{eqnarray}
where $\omega$, $j$ and $W$ are the energy, angular momentum and complex constant respectively. 
Using Eqs. (18), (19) and (20) in Eq. (17), a quadratic equation in 
$\frac{\partial R}{\partial r}$ is obtained as
 \begin{eqnarray}
  A\Bigg(\frac{\partial R}{\partial r}\Bigg)^2+B \left(\frac{\partial R}{\partial r}\right)+C=0,
  \end{eqnarray}
where the terms $A$, $B$ and $C$ are defined by 
\begin{eqnarray}
A&=&g^{rr}(1+2a_k)+2cm u^r u^r,\cr
B&=&4cm u^r u^\theta \left(\frac{\partial R}{\partial \theta}\right)-4cm u^t u^r (\omega-j\Omega),\cr
C&=&\left(g^{tt}(\omega-j\Omega)^2+g^{\theta \theta}\left(\frac{\partial R }{\partial \theta}\right)^2\right)(1+2a_k)+2cm u^t u^t(\omega-j\Omega)^2\cr && -4cm u^t u^\theta (\omega-j\Omega)\frac{\partial R }{\partial \theta}+2cm u^\theta u^\theta\left(\frac{\partial R }{\partial \theta}\right)^2+m^2.
\end{eqnarray}
   Then the two roots of the above equation are given by
\begin{eqnarray}
 R=\int \frac{-B\pm\sqrt{B^2-4AC}}{2A} dr. 
\end{eqnarray}
Completing the integral of Eq. (23) near the event horizon of KNdS black hole by applying residue theorem of complex analysis, the imaginary part of the two roots are given by
\begin{eqnarray}
R(r)_\pm=\frac{i\pi \Xi (r_h^2 +a^2)(\omega-j\Omega)[2cm c_t c_r\pm L]}{2(\beta^2 r_h-M-2\frac{\Lambda r_h^3}{3})[(1+2a_k)+2cm c_r^2]},
\end{eqnarray}
where $L=\sqrt{(1-2cmc_t^2 +2cm c_r^2)+4a_k (1+a_k-cmc_t^2+cmc_r^2)}$.  $R_+$ and $R_-$ are the outgoing and incoming wave respectively.
The probabilities which cross near the event horizon, $r=r_h$ of KNdS black hole are
\begin{eqnarray}
\Gamma_{emission}&=& exp(-2ImI)=exp[-2(Im S_+ + Im \delta)]
\end{eqnarray}
and
\begin{eqnarray}
\Gamma_{absorption}&= &exp(-2ImI)=exp[-2(Im S_- + Im \delta)].
\end{eqnarray}
The ingoing particle has the \text{100\%} change to enter the black hole in accordance with semiclassical approximation. This indicates that $Im  \delta = -Im S_-$. Since $S_-=-S_+$, the tunneling probability of outgoing particle is calculated as 
\begin{eqnarray}
\Gamma_{rate}&=&\frac{\Gamma_{emission}}{\Gamma_{absorption}}\cr
&=& exp[-2Im S_+ +2Im S_-],\cr
&=& exp[\frac{-2\pi \Xi \gamma _1(r_h^2+a^2)(\omega-j \Omega)}{(\beta^2 r_h-M-2\frac{ \Lambda r_h^3}{3})}],
\end{eqnarray}
where \begin{eqnarray}
\gamma_1=\frac{ \sqrt{1+4a_k+4a_k^2-(1+2a_k)(2c m c_t^2-2c m c_r^2)}}{1+2a_k+2cm c_r^2}.
\end{eqnarray}
The Hawking temperature near the event of KNdS black hole in Lorentz violation theory is given by
\begin{eqnarray}
T_h&=&\frac{1}{2\pi \gamma_1} \Big[\frac{r_h-M-\frac{2}{3}\Lambda r_h^3-\frac{1}{3}\Lambda r_ha^2}{\Xi (r_h^2+a^2)} \Big]\cr
&=&\gamma_1^{-1} T_H.
\end{eqnarray}

From the above equation, it is observed that the Hawking temperature is modified due to the presence of $\gamma_1$. The modified Hawking temperature may increase or decrease depending upon the choices of ether like vectors. In the absence of ether like vectors, the modified Hawking temperature, $T_1$ tends to original Hawking temperature, $T_H$ given in Eq. (8).
Similarly the heat capacity near the event horizon of black hole is given by
\begin{eqnarray}
C_h&=&\frac{2\pi \gamma_1\Xi (r_h^2+a^2)^2 (3r_h^2-3a^2-3\Lambda r_h^4-\Lambda a^2r_h^2-3q^2)}{3(a^4-r_h^4)+4a^2r_h^2(3-2\Lambda r_h^2)-\Lambda r_h^2(3r_h^4+a^4)+3q^2(a^2+3r_h^2)}\cr
&=&\gamma_1 C_H.
\end{eqnarray}
We also see that the heat capacity of KNdS black hole is modified due to the presence of extra term $\gamma_1$ in Eq. (30). This indicates that the modified heat capacity depends not only mass, charge, cosmological constant but also on ether like vectors. In the absence of Lorentz symmetry violation, Eq. (30) is consistent with original heat capacity given in Eq. (12).
Utilizing Eqs. (5) and (7) in Eq. (24), the imaginary part of outgoing particle near the event horizon of KNdS black hole can be written as
\begin{eqnarray}
Im S&= &\frac{\gamma_2}{2}\Big[\frac{\pi \Xi k_1^2k_2^2}{\alpha_1 \beta^2 k_1 k_2-\alpha_1^2 (M+A)}\omega+\frac{\pi \Xi a^2 \alpha_1}{\beta^2 k_1 k_2-\alpha_1(M+A)}\omega\cr&&
-\frac{\pi \Xi a\alpha_1 }{\beta^2 k_1 k_2-\alpha_1(M+A)}j\Big],
\end{eqnarray}
where 
\begin{eqnarray} 
 k_1&=&\left(1+\frac{4\Lambda M^2}{3\beta ^2\alpha_1}+\cdots\right),\cr
 k_2&=&\left(M+\sqrt{M^2-(a^2+q^2)\alpha_1}\right),\cr
 A&=&\frac{2\Lambda}{3 \alpha_1 ^3}\left(1+\frac{4\Lambda M^2}{3\beta ^2\alpha_1}+\cdots\right)^3\left(M+\sqrt{M^2-(a^2+q^2) \alpha_1}\right)^3,\cr
\gamma_{2}&=&\frac{ 2cmc_tc_r+\sqrt{1+4a_k+4a_k^2-(1+2a_k)(2c m c_t^2-2c m c_r^2)}}{1+2a_k+2cm c_r^2}.
\end{eqnarray}
For obtaining the maximum value of integration, we neglect the terms $O(M)^i$, $i\geq 2$ from the denominator of Eq. (30), where $M$ is the mass of KNdS black hole. Therefore Eq. (31) becomes
\begin{eqnarray}
Im S&=&\frac{\gamma_{2}}{2}\Big[\frac{\pi \Xi k_2^2}{\beta^2 \alpha_1(k_2-\frac{\alpha_1 M}{\beta^2})}\omega+\frac{\pi \Xi a^2 \alpha_1}{\beta^2 (k_2-\frac{\alpha_1 M}{\beta^2})}\omega-\frac{\pi \Xi a \alpha_1}{\beta^2 (k_2-\frac{\alpha_1 M}{\beta^2})}j\Big].
\end{eqnarray}
If a pair of virtual particle is created near the event horizon of KNdS black hole, the positive energy virtual particle tunnels out and materializes a real particle whereas the negative energy particle is absorbed by the black hole. This process reduces the mass and angular momentum of the KNdS black hole. We calculate the imaginary part of the action in the following integral form as 

\begin{align}
{\rm Im S}&=& \frac{\gamma_{2}}{2}\Big[\int^\omega_0\frac{\pi   \Xi k_2^2}{\beta^2 \alpha_1\sqrt{M^2-(a^2+q^2)\alpha_1} }d\omega^{'}+\int^\omega_0\frac{\pi \Xi a^2\alpha_1}{\beta^2\sqrt{M^2-(a^2+q^2)\alpha_1}}d\omega^{'}
 \cr&& -\int^j_0\frac{\pi \Xi a\alpha_1}{\beta^2\sqrt{M^2-(a^2+q^2)\alpha_1}}dj^{'}\Big].
\end{align} 
If the emitted particle is taken as an ellipsoid shell of energy $\omega$, then we fix ADM mass and angular momentum of the total space time. Hence it allows the KNdS black hole to vary. If a particle $\omega$ and angular momentum $j$ tunnel out from KNdS black hole, the mass $M$ and angular momentum $j$ can be replaced by $M-\omega$ and $J-j$ respectively. If we ignore the term $M(1-\frac{\alpha_1}{\beta^2})$ from Eq. (34), then it can be written as

 \begin{align}
  {\rm Im S}&=& \frac{\gamma_{2}}{2}\Bigg[\frac{-\pi\Xi }{\beta^2\alpha_1}\int^{M-\omega}_M\frac{  k_2^2}{\sqrt{(M-\omega)^2-(a^2+q^2)\alpha_1} }d(M-\omega^{'})\cr&&-\frac{\pi \Xi a^2\alpha_1}{\beta^2}\int^{M-\omega}_M\frac{1}{\sqrt{(M-\omega)^2-(a^2+q^2)\alpha_1}}d(M-\omega^{'})\cr&&
  +\frac{\pi \Xi a\alpha_1}{\beta^2}\int^{J-j}_j\frac{1}{\sqrt{(M-\omega)^2-(a^2+q^2)\alpha_1}}d(J-j)^{'}\Bigg],
\end{align} 
   where $J-j=(M-\omega^{'})a$.
   Then Eq. (35) can be written as
\begin{align} 
  {\rm Im S}&=\frac{-\gamma_{2}\pi\Xi }{2\beta^2\alpha_1 }\Bigg[\int^{M-\omega}_M\frac{2(M-\omega)^2+2(M-\omega)\sqrt{(M-\omega)^2-(a^2+q^2) \alpha_1}}{\sqrt{(M-\omega)^2-(a^2+q^2)\alpha_1}}d(M-\omega^{'})\cr&-\int^{M-\omega}_M \frac{(a^2+q^2)\alpha_1 }{\sqrt{(M-\omega)^2-(a^2+q^2) \alpha_1}}d(M-\omega^{'})\Bigg].
\end{align}
Completing the above integral, we get
\begin{align}
   {\rm Im S}&=\frac{-\gamma_{2}\pi\Xi }{2\beta^2\alpha_1 }\Big[(M-\omega)\sqrt{(M-\omega)^2-(a^2+q^2)\alpha_1}+(M-\omega)^2\cr&-M\sqrt{M^2-(a^2+q^2)\alpha_1}-M^2\Big]\cr
   &=\frac{-\gamma_{2}\pi\Xi }{4\beta^2\alpha_1}[(M-\omega)+\sqrt{(M-\omega)^2-(a^2+q^2)\alpha_1})^2-( M+\sqrt{M^2-(a^2+q^2)\alpha_1})^2]\cr
&=\frac{-\gamma_{2} \pi}{2}[\frac{(r_f^{2}+a^2)}{\Xi}-\frac{(r_i^{2}+a^2)}{\Xi}].
\end{align}
In accordance with WKB approximation, the tunneling rate of fermions is given by
\begin{eqnarray}
\Gamma= exp[-2Im S]=exp[\gamma_{2}\Delta S_{BH}],
\end{eqnarray}
where $\gamma_{2} \Delta S_{BH} =\gamma_{2} \Big[\frac{(r_f^2 +a^2)}{\Xi}-\frac{(r_i^2+a^2)}{\Xi}\Big]$  is the change in Bekenstein-Hawking entropy in Lorentz violation theory. $ r_{i}=\frac{\Xi}{\beta\sqrt{2}\alpha_1 }\Big[M+\sqrt{M^2-(a^2+q^2)\alpha_1}\Big]$ 

and  $ r_{f}=\frac{\Xi}{\beta\sqrt{2}\alpha_1}\Big[(M-\omega)+\sqrt{(M-\omega)^2-(a^2+q^2)\alpha_1}\Big]$ denote the locations of event horizon before and after emission of the particles with energy $\omega$ near the event horizon of KNdS black hole. From Eq. (38), we observe that the change in Bekenstein-Hawking entropy of KNdS black hole is modified due to presence of $\gamma_{2} $ given in Eq. (28). In the absence of Lorentz violation theory, the term $\gamma_{2} $ tends to unity and hence Eq. (38) will be original change in Bekenstein-Hawking entropy of KNdS black hole. This indicates that the change in Bekenstein-Hawking entropy depends on the ether like vectors $u^\alpha$.
  \section{Fermions tunneling in frame dragging coordinate}
  To investigate the tunneling of spin-$\frac{1}{2}$ fermions particle near the event horizon of KNdS black hole in four dimensional frame dragging coordinate, we take 
 $\phi=\varphi-\Omega t$ and $\Omega=-\frac{g_{14}}{g_{44}}$.
 Therefore Eq. (1) can be written as
  \begin{align}
 ds^2&=-\frac{\Delta_r\Delta_{\theta}\rho^{2}}{\Xi^{2}[\Delta_{\theta}(r^2+a^2)^2-\Delta_ra^2 \sin^2\theta]}dt^2+\frac{\Delta_\theta(r^2+a^2)^2-\Delta_r a^2\sin^2\theta}{\rho^2\Xi^2}\sin^2\theta d\phi^2\cr&+\frac{\rho^2}{\Delta_r}dr^2+\frac{\rho^2}{\Delta_{\theta}}d\theta^2.\cr&
\end{align}
The components of $\gamma^\mu$ matrices in frame dragging coordinates are choosen as
\begin{eqnarray}
\gamma^t&=&\frac{1}{\sqrt{g_{tt}}}
\begin{bmatrix}
i&o\\
0&-i\\
\end{bmatrix},\nonumber\\
\gamma^r&=&\sqrt{g_{rr}}
\begin{bmatrix}
0&\sigma^3\\
\sigma^3&0\\
\end{bmatrix},\nonumber\\
\gamma^\theta&=&\frac{1}{\sqrt{g_{\theta\theta}}}
\begin{bmatrix}
0&\sigma^1\\
\sigma^1&0\\
\end{bmatrix},\nonumber\\
\gamma^\phi&=&\frac{1}{\sqrt{g_{\phi\phi}}}
\begin{bmatrix}
0&\sigma^2\\
\sigma^2&0\\
\end{bmatrix},\end{eqnarray}
where $\sigma^j (j=1,2,3)'s$ are the Pauli Sigma matrices which satisfy anticommutation relation given in Eq. (14). The $\gamma^5$ matrix is calculated as
\begin{eqnarray}
\gamma^5=i\gamma^t \gamma^r \gamma^\theta \gamma^\phi=i\,   \Xi^2(\Delta_r \sin \theta)^{-1}
\begin{bmatrix}
0&-1\\
1&0\\
\end{bmatrix}.
\end{eqnarray}
To study the tunneling of fermions near the event horizon of KNdS black hole in frame dragging coordinate with Lorentz violation theory, the wave function $\psi$ is chosen as
\begin{eqnarray}
\psi=exp[\frac{i}{\hbar}I(t,r,\theta,\phi)].
\end{eqnarray}
The ether like vectors $u^\alpha (\alpha=1,2,3,4)$ from the KNdS black hole in 4-dimensional frame dragging coordinate are chosen as
 \begin{eqnarray}
   u^t&=&\frac{c_t}{\sqrt{-g_{tt}}} = \frac{c_t \Xi \sqrt{\Delta_{\theta}(r^2+a^2)^2-\Delta_r a^2 \sin^2\theta}}{\sqrt{\Delta_r\Delta_{\theta}\rho^{2}}} ,\cr
u^{r}&=&\frac{c_{r}}{\sqrt{g_{rr}}} =c_r\sqrt{\frac{\Delta_r}{\rho^2}},\cr
u^{\theta}&=&\frac{c_{\theta}}{\sqrt{g_{\theta\theta}}} =c_\theta \sqrt{\frac{\Delta_\theta}{\rho^2}},\cr
u^{\phi}&=&\frac{c_{\phi}}{\sqrt{g_{\phi\phi}}}=\frac{c_{\phi}\sqrt{\rho^2}\Xi}{\sqrt{[\Delta_{\theta}(r^2+a^2)^2-\Delta_r a^2 \sin^2\theta]\sin^2\theta}},
 \end{eqnarray}
 where $c_t, c_r, c_\theta$ and $c_\phi$ are arbitrary constants. It is observed that $u^\alpha (\alpha=1,2,3,4)$ are not constant from Eq. (43) but $u^\alpha u_\alpha $ satisfies the following condition
  \begin{eqnarray}
     u^{\alpha}u_{\alpha}=-c_t^2+c_r^2+c_\theta^2+c_\phi^2={\rm constant}.
\end{eqnarray}
 Using Eqs. $(39), (40), (42)$ and $(43)$ in Eq. $(13)$ and after some calculations, we obtain the dynamical equation of spin-$\frac{1}{2}$ fermions particle near the event horizon of KNdS black hole as

\begin{align}
&\left[ g^{tt}\left(\frac{\partial S}{\partial t}\right)^2+g^{rr}\left(\frac{\partial S}{\partial r}\right)^2+g^{\theta \theta}\left(\frac{\partial S}{\partial \theta}\right)^2+g^{\phi\phi}\left(\frac{\partial S}{\partial \phi}\right)^2\right](1+2a_k)+2cm u^t u^t\left(\frac{\partial S}{\partial t}\right)^2\cr &+4cm u^t u^r \left(\frac{\partial S}{\partial t}\right)\left(\frac{\partial S}{\partial r}\right) +4cm u^t u^\theta \left(\frac{\partial S}{\partial t}\right)\left(\frac{\partial S}{\partial \theta}\right)+4cm u^t u^\phi \left(\frac{\partial S}{\partial t}\right)\left(\frac{\partial S}{\partial \phi}\right)\cr &+2cm u^r u^r \left(\frac{\partial S}{\partial r}\right)^2+4cm u^r u^\theta \left(\frac{\partial S}{\partial r}\right)\left(\frac{\partial S}{\partial \theta}\right)+4cm u^r u^\phi \left(\frac{\partial S}{\partial r}\right)\left(\frac{\partial S}{\partial \phi}\right)\cr&+2cm u^\theta u^\theta \left(\frac{\partial S}{\partial r}\right)^2+4cm u^\theta u^\phi \left(\frac{\partial S}{\partial \theta}\right)\left(\frac{\partial S}{\partial \phi}\right)+2cm u^\phi u^\phi \left(\frac{\partial S}{\partial r}\right)^2+ m^2=0.
\end{align} 
We observe that Eq. (45) involves the variables $t, r, \theta$ and $\phi$. To investigate the tunneling of fermions near the event horizon of KNdS black hole in 4-dimensional frame dragging coordinate system, the action $S$ can be written as
\begin{eqnarray}
 S=-\omega t+H(r,\theta) +j \phi +Z,
\end{eqnarray}
where $\omega, j$ and $Z$ denote the energy, angular momentum and complex constant respectively.
Using Eq. (46) in Eq. (45), we obtain a quadratic equation in $\frac{\partial H}{\partial r}$ as
\begin{eqnarray}
  D\Bigg(\frac{\partial H}{\partial r}\Bigg)^2+E \left(\frac{\partial H}{\partial r}\right)+F=0,
  \end{eqnarray}
where the terms D, E and F are defined by
\begin{align}
D&=g^{rr}(1+2a_k)+2cm u^r u^r,\cr
E&=4cm u^r u^\theta \left(\frac{\partial H}{\partial \theta}\right)-4cm u^t u^r (\omega-j\Omega)+4cmu^r u^\phi j,\cr
F&=\left(g^{tt}(\omega-j\Omega)^2+g^{\theta \theta}\left(\frac{\partial H }{\partial \theta}\right)^2+g^{\phi\phi}j^2\right)(1+2a_k)+2cm u^t u^t(\omega-j\Omega)^2\cr & -4cm u^t u^\theta (\omega-j\Omega)\frac{\partial H }{\partial \theta}-4cm u^t u^\phi(\omega-j\Omega)j+4cm u^\theta u^\phi\left(\frac{\partial H }{\partial \theta}\right)j\cr&+2cm u^\theta u^\theta\left(\frac{\partial H }{\partial \theta}\right)^2+2cm u^\phi u^\phi j^2+m^2.
\end{align}
Since Eq. (47) is a quadratic equation, therefore the two roots are defined by
\begin{eqnarray}
 H=\int \frac{-E\pm\sqrt{E^2-4DF}}{2D} dr.
\end{eqnarray}
Completing the above integral by using residue theorem of complex analysis and Feynman prescription near the event horizon of KNdS black hole, the outgoing particle $(H_+)$ and ingoing particle $(H_-)$ are calculated as
\begin{eqnarray}
H(r)_\pm=\frac{i\pi \Xi (r_h^2 +a^2)(\omega-j\Omega)[2cm c_t c_r\pm L]}{2(\beta^2 r_h-M-2\frac{\Lambda r_h^3}{3})[1+2a_k+2cm c_r^2]},
\end{eqnarray} 

where $L=\sqrt{1+4a_k+4a_k^2-(1+2a_k)(2c m c_t^2-2c m c_r^2)}$.

The Hawking temperature of KNdS black hole in 4-dimensional frame dragging coordinate is derived as
\begin{eqnarray}
T_4&=&\frac{1}{2\pi \gamma_1} \Big[\frac{r_h-M-\frac{2}{3}\Lambda r_h^3-\frac{1}{3}\Lambda r_ha^2}{\Xi (r_h^2+a^2)} \Big]\cr
&=&\gamma_1^{-1} T_H.
\end{eqnarray}

We observe from Eqs. (24) and (50) that the radial action derived from near the event horizon of KNdS black hole in 3-dimensional frame dragging coordinate system is equal to the radial action obtained from 4-dimensional frame dragging coordinate system. Therefore the modified Hawking temperature, Bekenstein-Hawking entropy and heat capacity of 3-dimensional KNdS black hole are concordant with that of 4-dimensional KNdS black hole. The modified Hawking temperature, heat capacity and change in Bekenstein-Hawking near the event horizon of KNdS black hole in 4-dimensional frame dragging coordinate will be $\gamma_1^{-1} T_H ,\,\gamma_1 C_H$ and $\gamma_2 \Delta S_{BH}$ respectively. If $\gamma_1\in(1,\infty)$, the modified Hawking temperature $(T_h)$ is smaller than the original Hawking temperature $(T_H)$ but the modified heat capacity $(C_h)$ is bigger than the original heat capacity $(C_H)$. When $\gamma_1 \in (0,1)$, the modified Hawking temperature is bigger than the original Hawking temperature but the modified heat capacity is smaller than that of original heat capacity. The modified Hawking temperature and heat capacity are concordant with original Hawking temperature and heat capacity respectively if $\gamma_1 =1$. The change in Bekenstein-Hawking entropy  tends to the original change in Bekenstein-Hawking entropy when $\gamma_2=1$. The change in entropy increases ($\gamma_2 \Delta S_{BH}>\Delta S_{BH}$) or decreases ($\gamma_2 \Delta S_{BH}<\Delta S_{BH}$) if $\gamma_2\in(1,\infty)$ or $\gamma_2\in(0,1)$ respectively. This shows that the modified Hawking temperature, heat capacity and change in Bekenstein-Hawking entropy depend on ether like vectors $u^\alpha$.

  \section{Eddington coordinate} 
  
  To discuss modified Hawking temperature, change in Bekenstein-Hawking entropy and modified heat capacity near the event horizon of KNdS black hole with Lorentz symmetry violation in curved space time, we use a well-behaved coordinate system known as Eddington coordinate. 
  Let \begin{eqnarray}
  dt=du-\frac{\Xi}{\Delta_r}\sqrt{\frac{[\Delta_\theta(r^2+a^2) -\Delta_r a^2\sin^2 \theta]}{\Delta_\theta}}dr.
  \end{eqnarray}
  Using Eq. (52) in Eq. (6), we get
 \begin{eqnarray}
 ds^2=\frac{-\Delta_r \Delta_\theta \rho^2}{\Xi^2 K}du^2+\frac{2\rho^2}{\Xi}\sqrt{\frac{\Delta_\theta}{K}}dudr+\frac{\rho^2}{\Delta_\theta}d\theta^2.
\end{eqnarray} 
It is noted that all the metric components of the above equation are analytic near the event horizon of KNdS black hole. From Eq. (53), the ether like vectors $u^{\alpha} (\alpha=1,2,3)$ are chosen as
 
\begin{eqnarray}
u^u&=&\frac{c_u}{\sqrt{g_{11}}}=c_u\sqrt{\frac{-\Xi^2 K}{\Delta_r \Delta_\theta \rho^2}},\cr
u^r&=&\frac{c_r\sqrt{g_{11}}}{g_{12}}=\frac{c_r}{\rho^2}\sqrt{-\Delta_r \rho^2},\cr
u^{\theta}&=
&\frac{c_\theta}{\sqrt{g_{33}}}=c_\theta\sqrt{\frac{\Delta_\theta}{\rho^2}},
\end{eqnarray}
where $c_u, c_r $ and $c_\theta$ are arbitrary constants. Then the vectors $u^\alpha (\alpha=1,2,3)$ satisfy the following condition
\begin{eqnarray}
u^{\alpha}u_{\alpha}&=&c_u^2+2c_u c_r+c_{\theta}^2=\rm {constant}.
\end{eqnarray}  
From Eq. (13), we derive the dynamical equation of spin-$\frac{1}{2}$ fermions particle of KNdS black hole in Eddington coordinate as
\begin{align}
&\Big[g^{rr}\Big(\frac{\partial S}{\partial r}\Big)^2+g^{\theta\theta}\Big(\frac{\partial S}{\partial \theta}\Big)^2+2g^{ur}\Big(\frac{\partial S}{\partial u}\Big) \Big(\frac{\partial S}{\partial r}\Big) \Big](1+2a_k)\cr& +4cm u^u u^r \Big(\frac{\partial S}{\partial u}\Big)\Big(\frac{\partial S}{\partial r}\Big)+4cm u^u u^{\theta}\Big(\frac{\partial S}{\partial u}\Big)\Big(\frac{\partial S}{\partial \theta} \Big)+4cm u^r u^{\theta} \Big(\frac{\partial S}{\partial r}\Big) \Big(\frac{\partial S}{\partial \theta}\Big)\cr&+2cm u^u u^u \Big(\frac{\partial S}{\partial u}\Big)^2+2cm u^r u^r \Big(\frac{\partial S}{\partial r}\Big)^2+2cm u^{\theta} u^{\theta} \Big(\frac{\partial S}{\partial {\theta}}\Big)^2+m^2=0.
\end{align}
From Eq. (56), we know that the Hamilton principal function $S$ involves the partial derivatives with respect to $u,r$ and $\theta$. Since the Hawking radiation takes place along the radial direction, the variables can be separated as
\begin{eqnarray}
S=-\omega u+R^*(r)+\Theta(\theta)+j \phi+Z^*,
\end{eqnarray}
where $\omega, j$ and $Z^*$ denote the energy, angular momentum and complex constant respectively. Using Eq. (57) in Eq. (56), a quadratic equation in $\frac{\partial R^*}{\partial r} $ can be written as
\begin{eqnarray}
A_1\Big(\frac{\partial R^*}{\partial r}\Big)^2+ B_1\Big(\frac{\partial R^*}{\partial r}\Big)+C_1=0,
\end{eqnarray}  
where $A_1,B_1$ and $C_1$ are given by
 \begin{align}
 A_1&=\frac{\Delta_r}{\rho^2}{[(1+2a_k)-2cm c_r^2]},\cr
 B_1&=4cm c_r c_\theta \frac{\sqrt{-\Delta_r\Delta_\theta}}{\rho^2} \Big(\frac{\partial \Theta}{\partial \theta}\Big)-\frac{2\Xi}{\rho^2}\sqrt{\frac{K}{\Delta_ \theta}}(\omega-j\Omega)[(1+2a_k)+2cm c_u c_r],\cr
 C_1&=\frac{\Delta_\theta}{\rho^2}[(1+2a_k)+2cm c_\theta^2]\Big(\frac{\partial \Theta}{\partial \theta}\Big)^2 -2cm c_u^2 \frac{\Xi^2 k}{\Delta_r \Delta_\theta \rho^2}(\omega-j\Omega)^2 \cr&-4cm c_u c_\theta \frac{\Xi}{\rho^2}\sqrt{\frac{-K}{\Delta_r}}
(\omega-j\Omega)\Big(\frac{\partial \Theta}{\partial \theta}\Big)+m^2.
 \end{align}
 The two roots of Eq. (56) are calculated as
 \begin{eqnarray}
 R^*=\int \frac{-B_1 \pm \sqrt{B_1^2-4A_1C_1}}{2A_1}dr. 
 \end{eqnarray}
Eq. (60) has a singularity near the event horizon $r=r_h$ of KNdS black hole. To complete the integral by using Feynman prescription and residue theorem of complex analysis near the event horizon of black hole, we get
 \begin{eqnarray}
 R^*(r)_\pm &=&\frac{i\pi \Xi (r_h^2 +a^2)(\omega-j\Omega_h)[(1+2a_k+2cm c_u c_r)\pm A_2]}{2(\beta^2 r_h-M-2\frac{\Lambda r_h^3}{3})[(1+2a_k)-2cm c_r^2]},
 \end{eqnarray}
 where $A_2=\sqrt{(1+2a_k)[(1+2a_k)+4cm c_u c_r +2cm c_u^2 ]}$. $R^*_+$ and $R^*_-$ are the outgoing particle and ingoing particle respectively. Since the black hole evaporation is a quantum tunneling process [2], the probabilities of emission and absorption crossing near the event horizon of $r=r_h$ are calculated as [22]
 \begin{eqnarray}
\Gamma_{emission}&=& exp(-2ImI)=exp[-2(Im S_+ + Im \delta)]
\end{eqnarray} and
\begin{eqnarray}
\Gamma_{absorption}&= &exp(-2ImI)=exp[-2(Im S_- + Im \delta)].
\end{eqnarray}
The tunneling rate of the fermions particle near the event horizon of KNdS black hole in Eddington coordinate is calculated as
\begin{eqnarray}
\Gamma_{rate}&=&\frac{\Gamma_{emission}}{\Gamma_{absorption}}\cr
&=& exp[-2Im S_+ +2Im S_-]\cr
&=& exp[\frac{-2\pi \Xi \gamma_5 (r_h^2+a^2)(\omega-j \Omega)}{(\beta^2 r_h-M-2\frac{ \Lambda r_h^3}{3})}], 
 \end{eqnarray}
 where
 \begin{eqnarray}
\gamma_5=\frac{ \sqrt{(1+2a_k)[(1+2a_k)+4cm c_u c_r +2cm c_u^2 ]}}{1+2a_k-2cm c_r^2}. 
 \end{eqnarray}
 The Hawking temperature near the event horizon of KNdS black hole in Eddington coordinate with Lorentz violation theory is derived as
\begin{eqnarray}
T_U&=&\frac{(1-\frac{\Lambda a^2}{3}) r_h-M-\frac{2\Lambda r_h^3}{3}}{2\pi \Xi\gamma_5(r_h^2+a^2)}\cr
&=& \gamma_5^{-1} T_H.
\end{eqnarray}
It is observed from Eq. (66) that the Hawking temperature of KNdS black hole is modified due to the presence of arbitrary constant $\gamma_5$. If $\gamma_5$ tends to unity, Eq. (66) approaches to the original Hawking temperature ($T_H$) given in Eq. (8).
The modified Hawking temperature ($T_U$) of KNdS black hole would increase ($T_H<T_U$) or decrease ($T_H>T_U$) if $0<\gamma_5 <1$ or $1<\gamma_5 <\infty$ respectively. This indicates that the modified Hawking temperature of KNdS black hole in Eddington coordinate depends on the ether like vectors $u^\alpha$.
 The corresponding heat capacity near the event horizon of KNdS black hole is calculated as
 \begin{eqnarray}
C_{U}&=&\frac{2\pi \gamma_5\Xi (r_h^2+a^2)^2 (3r_h^2-3a^2-3\Lambda r_h^4-\Lambda a^2r_h^2-3q^2)}{3(a^4-r_h^4)+4a^2r_h^2(3-2\Lambda r_h^2)-\Lambda r_h^2(3r_h^4+a^4)+3q^2(a^2+3r_h^2)}\cr
&=& \gamma_5 C_H.
\end{eqnarray}
The heat capacity of KNdS black hole is also modified due to presence of $\gamma_5$ in the above equation. If $0<\gamma_5 <1$ or $1<\gamma_5 <\infty$, the modified heat capacity of black hole decreases $(C_U<C_H)$ or increases $(C_U>C_H)$ respectively. If $\gamma_5$ is equal to unity, then Eq. (67) is consistent with the original heat capacity $(C_U=C_H)$ given in Eq. (12). 
Substituting Eqs. (5) and (7) in Eq. (60), the imaginary part of the outgoing particle can be written as
\begin{align}
Im R(r)&= \frac{\gamma_6}{2}\Big[\frac{\pi \Xi k_1^2k_2^2}{\alpha_1 \beta^2 k_1 k_2-\alpha_1^2 (M+A)}\omega+\frac{\pi \Xi a^2 \alpha_1}{\beta^2 k_1 k_2-\alpha_1(M+A)}\omega\cr&
-\frac{\pi \Xi a\alpha_1 }{\beta^2 k_1 k_2-\alpha_1(M+A)}j\Big],
\end{align}
where 
\begin{align} 
 k_1&=\left(1+\frac{4\Lambda M^2}{3\beta ^2\alpha_1}+\cdots\right),\cr
 k_2&=\left(M+\sqrt{M^2-(a^2+q^2)\alpha_1}\right),\cr
 A&=\frac{2\Lambda}{3 \alpha_1 ^3}\left(1+\frac{4\Lambda M^2}{3\beta ^2\alpha_1}+\cdots\right)^3\left(M+\sqrt{M^2-(a^2+q^2) \alpha_1}\right)^3,\cr
\gamma_6&=\frac{(1+2a_k+2cmc_uc_r)+\sqrt{(1+2a_k)[(1+2a_k)+4cm c_uc_r+2cm c_u^2]}}{1+2a_k-2cm c_r^2}.
\end{align}

Evaluating the similar manner as done in Eqs. (33-37),  the tunneling rate of fermions in accordance with WKB approximation is given by
\begin{eqnarray}
\Gamma= exp[-2Im S]=exp[\gamma_{6}\Delta S_{BH}],
\end{eqnarray}
where
 $\gamma_{6} \Delta S_{BH} =\gamma_{6} \Big[\frac{(r_f^2 +a^2)}{\Xi}-\frac{(r_i^2+a^2)}{\Xi}\Big]$  is the change in Bekenstein-Hawking entropy in Lorentz violation theory. 
$r_i$ and $r_f$ are the locations of event horizon before and after the emission of particles with energy $\omega$ given in Eq. (38). The presence of $\gamma_6$ in Eq. (70) indicates that the change in Bekenstein-Hawking entropy is modified in Lorentz violation theory. The change in Bekenstein-Hawking entropy given above increases ($\gamma_{6}\Delta S_{BH}>\Delta S_{BH}$) or decreases ($\gamma_{6}\Delta S_{BH}<\Delta S_{BH}$)  near the event horizon of KNdS black hole if $0<\gamma_6 <1 $ or $1<\gamma_6<\infty$.
 When $\gamma_6$ tends to unity, Eq. (70) would be the original change in Bekenstein-Hawking entropy near the event horizon of KNdS black hole.
\section{Painleve coordinate}
To find the line element of Painleve coordinate system in KNdS black hole, we perform the transformation as
\begin{eqnarray}
dt=dT+F(r,\theta)dr+G(r,\theta)d\theta,
\end{eqnarray}
where $F(r,\theta)$ and $G(r,\theta)$ represent the two arbitrary functions satisfying the following condition
\begin{eqnarray}
\frac{\partial F(r,\theta)}{\partial\theta}=\frac{\partial G(r,\theta)}{\partial r}.
\end{eqnarray}
It is obvious that  the constant-time slice is a flat Euclidean space and taking
\begin{eqnarray}
g_{22}+F^2(r,\theta)\hat{g}_{11}=1,
\end{eqnarray}
where
\begin{eqnarray}
\hat{g}_{11}=-\frac{\Delta_r\Delta_\theta\rho^2}{\Xi^2[\Delta_\theta(r^2+a^2)^2-\Delta_r a^2\sin^2\theta]}.
\end{eqnarray}
After doing the above transformation, the line element Eq. (6) will be
\begin{align}
ds^2&=\hat{g}_{11}dT^2+2\sqrt{\hat{g}_{11}\Big(1-\frac{\rho^2}{\Delta_r}\Big)}dTdr+dr^2
+\Big(\frac{\rho^2}{\Delta_\theta}+G^2\hat{g}_{11}\Big)d\theta^2\cr&+2\hat{g}_{11}GdT d\theta+2G\sqrt{\hat{g}_{11}\Big(1-\frac{\rho^2}{\Delta_r}\Big)}dr d\theta.
\end{align}

The above line element has many attractive features. Firstly the components of the metric are analytic at the radius of the event horizon. Secondly there exists a line like killing vector field that makes the space time stationary. Thirdly the constant-time slices are flat Euclidean space which is very important because WKB approximation can be utilized in finding the emission rate.
The line element of Eq. (75) satisfies the Landau coordinate clockwise synchronization condition.

The contravariant metric components $g^{ab}$ of the above line element from Eq. (75) are found as
\begin{eqnarray}
g^{11}&=&\frac{1}{\hat{g}_{11}\rho^2}[\Delta_r+\Delta_\theta G^2\hat{g}_{11}],\,\,\,\,\,\,g^{23}=g^{32}=0,\,\,\,\,\,\,g^{22}=\frac{\Delta_r}{\rho^2},\cr
g^{12}&=&-\frac{\Delta_r}{\hat{g}_{11}\rho^2}\sqrt{\hat{g}_{11}\Big(1-\frac{\rho^2}{\Delta_r}\Big)},\,\,\,\,\,\,g^{31}=-\frac{G\Delta_\theta}{\rho^2},\,\,\,\,\,\,g^{33}=\frac{\Delta_\theta}{\rho^2}.
\end{eqnarray}
The ether like vectors $u^{\alpha}$ can be constructed from Eq. (75) as follows
\begin{eqnarray}
u^T&=&\frac{c_{T}}{\sqrt{-\hat{g}_{11}}},\cr
u^r&=&\frac{c_{r}\sqrt{-\hat{g}_{11}}}{g_{12}}=\frac{c_r\sqrt{\Delta_r}}{\sqrt{(\rho^2-\Delta_r)}},\cr
u^{\theta}&=&\frac{c_{\theta}}{\sqrt{g_{33}}}=\frac{c_{\theta}}{\sqrt{\frac{\rho^{2}}{\Delta_{\theta}}+G^{2}\hat{g}_{11}}},
\end{eqnarray}
where $c_{T}$, $c_r$ and $c_{\theta}$ are arbitrary constants and  $u^{\alpha}u_{\alpha}=-c^2_{T}+2 c_Tc_r+c^2_{\theta}={\rm constant}$ (near the event horizon of black hole). Substituting  Eqs. $(75)$ and $(77)$ in Eq. (13), we get
\begin{align}
&\Big[g^{TT}\Big(\frac{\partial S}{\partial T}\Big)^2+g^{rr}\Big(\frac{\partial S}{\partial r}\Big)^2+g^{\theta\theta}\Big(\frac{\partial S}{\partial \theta}\Big)^2+2g^{Tr}\Big(\frac{\partial S}{\partial T}\Big) \Big(\frac{\partial S}{\partial r}\Big) +2g^{T\theta}\Big(\frac{\partial S}{\partial T}\Big) \Big(\frac{\partial S}{\partial \theta}\Big)\Big](1+2a_k)\cr& +4cm u^T u^r \Big(\frac{\partial S}{\partial u}\Big)\Big(\frac{\partial S}{\partial r}\Big)+4cm u^T u^{\theta}\Big(\frac{\partial S}{\partial T}\Big)\Big(\frac{\partial S}{\partial \theta} \Big)+4cm u^r u^{\theta} \Big(\frac{\partial S}{\partial r}\Big) \Big(\frac{\partial S}{\partial \theta}\Big)\cr&+2cm u^T u^T \Big(\frac{\partial S}{\partial T}\Big)^2+2cm u^r u^r \Big(\frac{\partial S}{\partial r}\Big)^2+2cm u^{\theta} u^{\theta} \Big(\frac{\partial S}{\partial {\theta}}\Big)^2+m^2=0.
\end{align}
It is difficult to solve the above Eq. (78) because it contains the three independent variables $T, r$ and $\theta$. To find the radial action near the event horizon of KNdS black hole, the Hamilton principal function $S$ can be written as
\begin{eqnarray}
S=-\omega T+Z(r,\theta)+j \phi .
 \end{eqnarray}
 where $\omega$ and $j$ denote the energy of the particle and angular momentum respectively. Substituting Eq. (79) in Eq. (78), a quadratic equation in 
 $\frac{\partial Z}{\partial r}$ can be written as
\begin{eqnarray}
D_1\Big(\frac{\partial Z}{\partial r}\Big)^2+E_1\Big(\frac{\partial Z}{\partial r}\Big)+F_1=0,
\end{eqnarray}
where $D_1$, $E_1$ and $F_1$ are defined by
\begin{eqnarray}
D_1&=& \Delta_r\Big[\frac{(1+2a_k)}{\rho^2}+\frac{2cmc_r ^2}{\rho^2-\Delta_r}\Big],\cr
E_1&=&\frac{[(-2\Delta_r +2\rho^2)(1+2a_k)+4cmc_T c_r \rho^2]\Xi \sqrt{K}(-\omega+j\Omega)}{\rho^2 \sqrt{\Delta_\theta \rho^2 (\rho^2-\Delta_r)}}\cr&&+4cm c_r c_\theta \frac{\sqrt{\Delta_r \Delta_\theta}}{\sqrt{(\rho^2-\Delta_r)(\hat{g_{11}}G^2 \Delta_\theta+\rho^2)}},\cr
F_1&=&\frac{\Delta_r}{\rho^2 \hat{g_{11}}} (-\omega+j\Omega)^2(1+2a_k)+\frac{\Delta_\theta}{\rho^2}[G(-\omega+j\Omega)-(\frac{\partial Z}{\partial \theta})]^2(1+2a_k)\cr&&-\frac{2cm c_T^2}{\hat{g_{11}}}(-\omega+j\Omega)^2+\frac{4cm c_T c_\theta}{\sqrt{-{\hat{g_{11}}({\hat{g_{11}} G^2+\frac{\rho^2}{\Delta_\theta})}}}}(-\omega+j\Omega)\Big(\frac{\partial Z}{\partial \theta}\Big)
\cr&&+\frac{2cm c_\theta^2 \Delta_\theta}{\Delta_\theta \hat{g_{11}}G^2+\rho^2}\Big(\frac{\partial Z}{\partial \theta}\Big)^2+m^2.
\end{eqnarray}
The two roots of Eq. (80) are calculated as
\begin{eqnarray}
 Z(r)=\int \frac{-E_1 \pm \sqrt{E_1^2-4D_1F_1}}{2D_1}dr .
\end{eqnarray}
It is noted that Eq. (82) has a singularity near the event horizon. We calculated the above integral by applying Feynman prescription and residue theorem of complex analysis as
\begin{eqnarray}
Z(r)_\pm &=& \frac{i\pi \Xi(r^2+a^2)(\omega-j\Omega)[{(1+2a_k)+2cm c_T c_r}\pm A_2]}{2(\beta^2 r_h-M-\frac{2r_h^3}{l^2}){[(1+2a_k)+2cm c_r^2}]},
\end{eqnarray}

where $Z_+ (r)$ and $Z_- (r)$ denote the outgoing wave and ingoing wave respectively and $A_2=\sqrt{(1+2a_k)^2+4(1+2a_k)cm c_T c_r-(1+2a_k)2cm c_T^2}$.

The tunneling probability near the event horizon of KNdS black hole in Painleve coordinate is obtained as
\begin{eqnarray}
\Gamma_{rate}&=&\frac{\Gamma_{emission}}{\Gamma_{absorption}}=exp[-2Im S_++2Im S_-]\cr &=&\frac{-2\pi \gamma_7 \Xi(r^2_h+a^2)(\omega-j\Omega)}{(\beta^2 r_h-M-\frac{2r_h^3}{l^2})}{
},
\end{eqnarray}
which is similar to the Boltzmann factor; exp(-$\omega\beta_0$), where $\beta_0$ is the inverse temperature of KNdS black hole. Then Hawking temperature is derived as
\begin{align}
T_p&=\frac{(1-\frac{\Lambda a^2}{3}) r_h-M-\frac{2\Lambda r_h^3}{3}}{2\pi \Xi\gamma_7(r_h^2+a^2)},
\end{align}
where $\gamma_7=\frac{\sqrt{(1+2a_k)[(1+2a_k)+4cm c_T c_r-2cm c_T^2]}}{{1+2a_k+2cm c_r^2}}$. From Eq. (85), the heat capacity is calculated as
\begin{align}
C_{hp}&=&\frac{2\pi \gamma_7\Xi (r_h^2+a^2)^2 (3r_h^2-3a^2-3\Lambda r_h^4-\Lambda a^2r_h^2-3q^2)}{3(a^4-r_h^4)+4a^2r_h^2(3-2\Lambda r_h^2)-\Lambda r_h^2(3r_h^4+a^4)+3q^2(a^2+3r_h^2)}.
\end{align}
From Eqs. (85) and (86), it is noted that the Hawking temperature and heat capacity of KNdS black hole are modified due to the presence of arbitrary constant $\gamma_7$, which appears from the Lorentz symmetry violation theory. If $0<\gamma_7<1$, the modified Hawking temperature increases but heat capacity decreases near the event horizon of KNdS black hole. Again, when $1<\gamma_7<\infty$, the modified Hawking temperature decreases but heat capacity increases of KNdS black hole. If Lorentz violation is cancelled, then $\gamma_7$ tends to unity. In such case, Eqs. (85) and (86) are consistent with Eqs. (8) and (12) respectively.

To investigate the entropy of black hole, substituting Eqs. (5) and (7) in Eq. (83), the imaginary part of the outgoing action is given by
\begin{eqnarray}
Im S&= &\frac{\gamma_8}{2}\Big[\frac{\pi \Xi k_1^2k_2^2}{\alpha_1 \beta^2 k_1 k_2-\alpha_1^2 (M+A)}\omega+\frac{\pi \Xi a^2 \alpha_1}{\beta^2 k_1 k_2-\alpha_1(M+A)}\omega\cr&&
-\frac{\pi \Xi a\alpha_1 }{\beta^2 k_1 k_2-\alpha_1(M+A)}j\Big],
\end{eqnarray}
where 
\begin{align} 
 k_1&=\left(1+\frac{4\Lambda M^2}{3\beta ^2\alpha_1}+\cdots\right),\cr
 k_2&=\left(M+\sqrt{M^2-(a^2+q^2)\alpha_1}\right),\cr
 A&=\frac{2\Lambda}{3 \alpha_1 ^3}\left(1+\frac{4\Lambda M^2}{3\beta ^2\alpha_1}+\cdots\right)^3\left(M+\sqrt{M^2-(a^2+q^2) \alpha_1}\right)^3,\cr
 \gamma_8 &=\frac{[{(1+2a_k)+2cm c_T c_r}+\sqrt{(1+2a_k)[(1+2a_k)+2cm c_T (2c_r- c_T)}]}{
{1+2a_k+2cm c_r^2}}.
\end{align}
As calculated in Eqs. (33-37), the tunneling rate of fermions in accordance with WKB approximation is given by
\begin{eqnarray}
\Gamma= exp[-2Im S]=exp[\gamma_{8}\Delta S_{BH}],
\end{eqnarray}

where $\gamma_{8} \Delta S_{BH} =\gamma_{8} \Big[\frac{(r_f^2 +a^2)}{\Xi}-\frac{(r_i^2+a^2)}{\Xi}\Big]$  is the change in Bekenstein-Hawking entropy in Lorentz violation theory. $r_f$ and $r_i$ denote the locations of event horizon after and before the emission of particles of energy $\omega$ derived in Eq. (38).

The change in Bekenstein-Hawking entropy may increase or decrease depending upon the values of $\gamma_8 \in (1,\infty)$ or $\gamma_8 \in (0,\infty)$ respectively. If $\gamma_8=1$, Eq. (89) will be original change in Bekenstein-Hawking entropy of KNdS black hole.
This shows that the Hawking temperature, heat capacity and change in Bekenstein-Hawking entropy near the event horizon of KNdS black hole are affected due to Lorentz violation theory.

\section{Discussion and Conclusion}
The original Hawking temperature ($T_H$) and  modified Hawking temperatures of KNdS black hole derived from dragging coordinate systems ($T_h$), Eddington coordinate system ($T_U$) and Painleve coordinate system ($T_{p}$) by using Dirac equation with Lorentz violation theory in curved space time can be combined to a single equation having different constant terms as follow
\begin{eqnarray}
T&=&\frac{1}{\delta}\times  \frac{[r_h - M-\frac{2\Lambda r_h^3}{3}-\frac{\Lambda  r_h a^2}{3}]}{2\pi \Xi(r_h^2+a^2)}.
\end{eqnarray}

Case (i) T tends to the modified Hawking temperature ($T_h$) of KNdS black hole derived from frame dragging coordinates given in Eqs. (29) and (51) if
\begin{eqnarray}
 \delta &=&\frac{\sqrt{1+4a_k+4a_k^2-(1+2a_k)(2c m c_t^2-2c m c_r^2)}}{1+2a_k +2cm c_r^2}.
\end{eqnarray}
When $\delta=1$, the effect of Lorentz violation theory is cancelled. Therefore Eq. (90) becomes the original Hawking temperature $(T_H)$ near the event horizon of KNdS black hole given in Eq. (8). If $\delta \in (0,1)$ or $\delta \in (1,\infty)$, the modified Hawking temperature will increase $(T_H<T=T_h)$ or decrease $(T_H>T=T_h)$ respectively.

Case (ii) T approaches to the modified Hawking temperature $(T_U)$ near the event horizon of KNdS black hole obtained from Eddington coordinate given in Eq. (66) if
\begin{eqnarray}
 \delta &=&\frac{\sqrt{(1+2a_k)[(1+2a_k)+4c m c_u c_r +2c m c_u^2]}}{1+2a_k-2cm c_r^2}.
\end{eqnarray}
If $0<\delta<1$, the modified Hawking temperature $(T=T_U)$ is less than original Hawking temperature $(T_H)$ (i.e., $T=T_U<T_H$). When $1<\delta<\infty$, the modified Hawking temperature ($T=T_U$) is greater than original Hawking temperature $(T_H)$ (i.e., $T=T_U>T_H).$ If there is no Lorentz violation in Eq. (90), then $T=T_U=T_H$ and $\delta$ tends to unity.

Case (iii) Eq. (90) tends to the modified Hawking temperature $(T_p)$ of KNdS black hole calculated from well-behaved Painleve coordinate given in Eq. (85) if
\begin{eqnarray}
 \delta&=&\frac{\sqrt{(1+2a_k)[(1+2a_k)+4c m c_T c_r +2c m c_T^2]}}{1+2a_k+2cm c_r^2}.
\end{eqnarray}
The modified Hawking temperature increases ($T_H<T=T_p$) or decreases ($T_H>T=T_p$) according to $\delta\in(0,1)$  or  $\delta\in(1,\infty)$ respectively. If $\delta$ approaches to unity, the modified Hawking temperature $(T=T_p)$ is equal to original Hawking temperature given in Eq. (8) $(T_H=T=T_p)$. The variation of Hawking temperature near the event horizon of KNdS black hole is shown graphically in Fig. [1] by taking the parameters: $a=0.2, \Lambda=0.6, c=0.8, m=0.5, c_t=0.5, c_r=0.3, c_T=1.3, c_u=0.99, a_k=0.3, q=0.1$. For the above set of parameters, the original Hawking temperature $(T_H)$, modified Hawking temperatures obtained from dragging coordinate $(T_h)$, well behaved Eddington coordinate $(T_U)$ and Painleve coordinate $(T_p)$ are zero when $r_h=0.228126$ and $r_h=1.26542$. If $0.228126<r_h<1.26542$, the Hawking temperatures obey the inequality, $T_p<T_h<T_H<T_e$. If $1.26542<r_h<\infty$, the Hawking temperatures satisfy the inequality, $T_e<T_H<T_h<T_p$. From the above set of parameters, the maximum value of the original Hawking temperature $(T_H)$, modified Hawking temperatures derived from dragging coordinate $(T_h)$, Eddington coordinate $(T_U)$ and Painleve coordinate $(T_p)$ are $0.0923708, 0.100637, 0.0659887$ and $0.130753$ respectively when $r_h=0.415224$.

The original heat capacity (12) and modified heat capacities near the event horizon of KNdS black hole  obtained from frame dragging coordinates (30), Eddington coordinate (67) and Painleve coordinate (86) can be combined to a single equation with different constant terms as
\begin{equation}
C_{D}=\tau\frac{2\pi \Xi (r_h^2+a^2)^2 (3r_h^2-3a^2-3\Lambda r_h^4-\Lambda a^2r_h^2-3q^2)}{3(a^4-r_h^4)+4a^2r_h^2(3-2\Lambda r_h^2)-\Lambda r_h^2(3r_h^4+a^4)+3q^2(a^2+3r_h^2)}.
\end{equation}
Case (i) If 
\begin{eqnarray}
\tau &=& \frac{ \sqrt{(1-2cmc_t^2 +2cm c_r^2)+4a_k (1+a_k-cmc_t^2+cmc_r^2)}}{1+2a_k+2cm c_r^2},
\end{eqnarray}
Eq. (94) represents the modified heat capacity given in Eq. (30). When $\tau=1$, the effect of Lorentz violation theory is cancelled. In such case the modified heat capacity $(C_D)$ is equal to original heat capacity $(C_H)$ given in Eq. (12). If $0<\tau<1$ or $1<\tau<\infty$, the modified heat capacity decreases ($C_D=C_h<C_H$) or increases  ($C_D=C_h>C_H$) near the event horizon of black hole.
\begin{figure}
\centering
\includegraphics{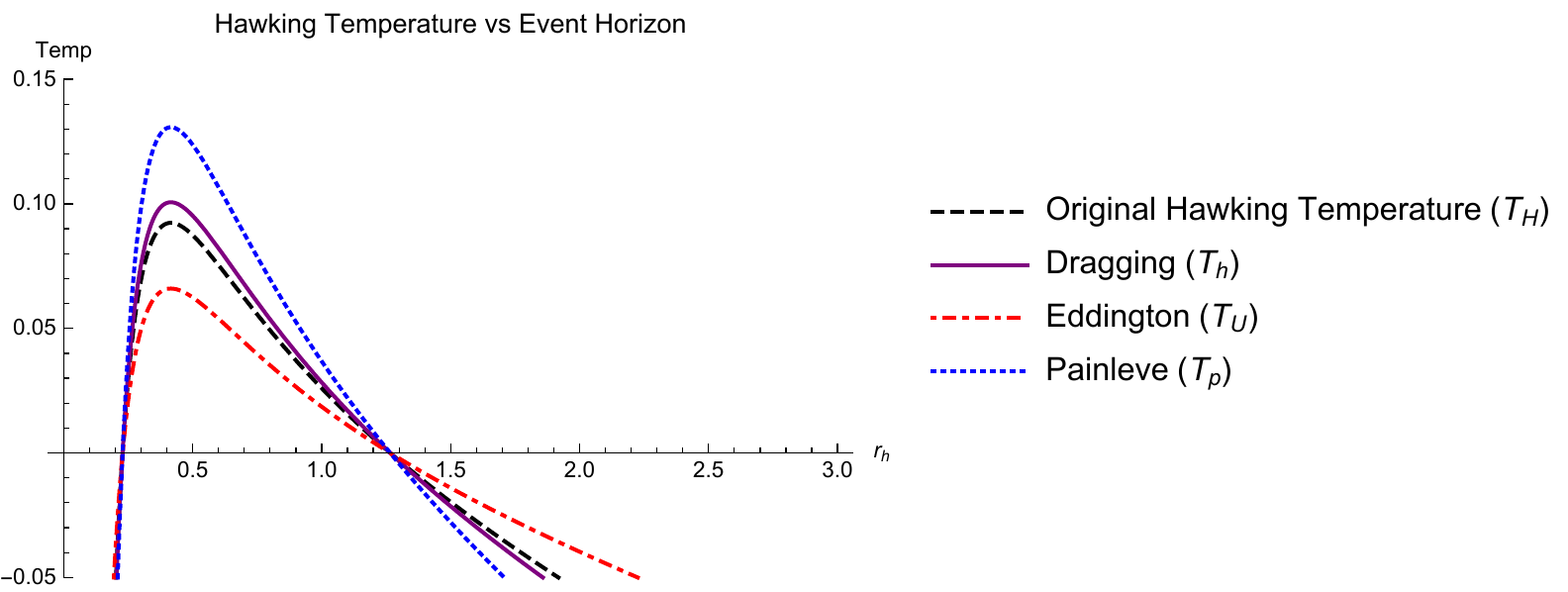}
\caption{Plot of original and modified Hawking temperatures of KNdS black hole derived from different coordinate systems- Frame dragging coordinates,  Eddington coordinate  and Painleve coordinate with radius of event horizon, ${r_h}$.}
\end{figure}

Case (ii) Eq. (94) is the modified heat capacity given in Eq. (67) if
\begin{eqnarray}
\tau &=& \frac{ \sqrt{(1+2a_k)^2+4cm c_u c_r (1+2a_k)+2cm c_u^2 (1+2a_k)}}{1+2a_k-2cm c_r^2}.
\end{eqnarray}
When $\tau=1$, the Lorentz violation becomes zero and Eq. (94) is equal to original heat capacity given in Eq. (12). The modified heat capacity increases ($C_D=C_U>C_H$) if $\tau\in(1,\infty)$ and the modified heat capacity decreases ($C_D=C_U<C_H$) if $\tau\in(0,1)$. If $q=0$ in Eq. (94), then the similar  finding is obtained in [55]. If we set $\Lambda=0$ in Eq. (94), the similar result in the tunneling of scalar particle is discussed in [56].

Case (iii) If
\begin{eqnarray}
\tau &=& \frac{\sqrt{(1+2a_k)^2+4(1+2a_k)cm c_T c_r-(1+2a_k)2cm c_T^2}}{{1+2a_k+2cm c_r^2}},
\end{eqnarray}
Eq. (94) is the modified heat capacity near the event horizon of KNdS black hole given in Eq. (86). The modified heat capacity decreases ($C_D=C_{hp}<C_H$) if $0<\tau<1$ and modified heat capacity increases ($C_D=C_{hp}>C_H$) if $1<\tau<\infty$. If $\tau=1$, the Lorentz violation has been cancelled. In such case the modified heat capacity, $C_D$ is equal to original heat capacity, $C_H$ derived in Eq. (12) ($C_D=C_{hp}=C_H$). The modified heat capacity of KNdS black hole is shown graphically with the radius of event horizon in Fig. [2]. For the set of parameters;  $a=0.3$, $\Lambda=0.06$, $c=0.5$, $m=1$, $c_t=1.1$, $c_r=0.5$, $c_T=1.9$, $c_u=0.99$, $a_k=0.4$, $q=0.1$, the KNdS black hole has a phase transition when $r_h=0.630317$. The dotted green line indicates the position of phase transition of KNdS black hole for the above parameters. Similar phase transition of KNdS black hole was observed in [64]. If $0<r_h<0.630317$, the original heat capacity $(C_H)$, modified heat capacities derived from dragging coordinate $(C_h)$, Eddington coordinate $(C_U)$ and Painleve coordinate $(C_{hp})$  satisfy the inequality $C_U<C_H<C_h<C_{hp}$. If $0.630317<r_h<\infty$, the heat capacities obey the inequality $C_{hp}<C_h<C_H<C_U$.

If $a=\Lambda=0$ in Eq. (94), then it can be written as
\begin{eqnarray}
C_{RH} =\frac{2\pi\tau_1 (\sqrt{M^2-Q^2)}(2M^2-Q^2+2M \sqrt{M^2-Q^2})}{(M-2\sqrt{M^2-Q^2})},
\end{eqnarray}
where $\tau_1$ is an arbitrary constant derived from ether like vectors $u^{\alpha}$.

Eq. (98) is the heat capacity of Reissner-Nordstrom black hole with Lorentz violation theory. The heat capacity of Eq. (98) has infinite point of discontinuity at $\mid Q\mid/M = \sqrt{3}/2$.  It is also noted that the Reissner-Nordstrom black hole has a phase transition at the charge to mass ratio of $\sqrt{3}/2$. We observe that the position of phase transition is not affected by Lorentz violation theory.  In the absence of Lorentz violation theory, Eq. (98) is consistent with earlier paper [65].

\begin{figure}
\centering
\includegraphics{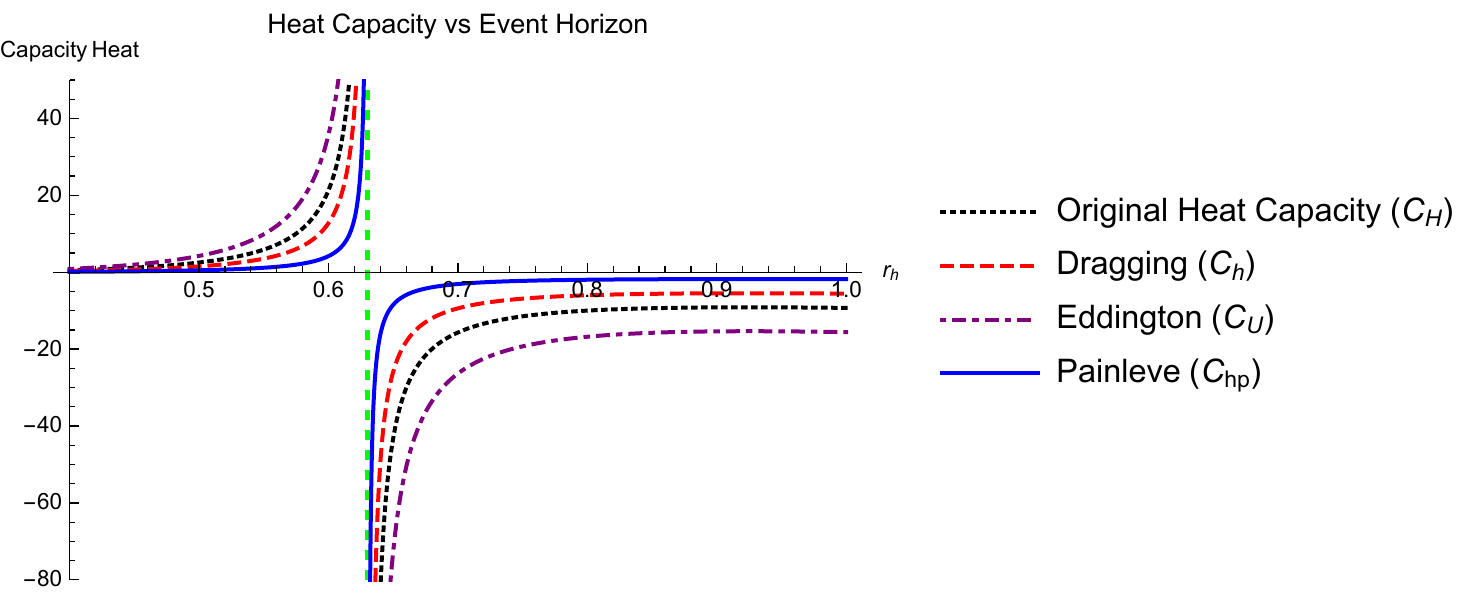}
\caption{Plot of modified heat capacity, $C_h$ with radius of event horizon, ${r_h}$ of KNdS black hole for different coordinate systems- Frame dragging coordinate, Painleve and Eddington. The original heat capacity $C_H$ is also shown.}
\end{figure}

The three expressions of change in Bekenstein-Hawking entropies of KNdS black hole obtained from dragging coordinates, Eddington coordinate and Painleve coordinate can be recast to a single equation as follows
\begin{eqnarray}
\eta \Delta S_{BH} =\eta \Big[\frac{(r_f^2 +a^2)}{\Xi}-\frac{(r_i^2+a^2)}{\Xi}\Big].
\end{eqnarray} 

Case (i) Eq. (99) is the change in Bekenstein-Hawking entropy given in Eq. (38) if
\begin{equation}
 \eta=\frac{ 2cmc_tc_r+\sqrt{(1-2cmc_t^2 +2cm c_r^2)+4a_k (1+a_k-cmc_t^2+cmc_r^2)}}{[(1+2a_k)+2cm c_r^2]}.
\end{equation}
If $\eta=1$, the Lorentz violation is cancelled and Eq. (99) is the original change of Bekenstein-Hawking entropy of KNdS black hole. The change in Bekenstein-Hawking entropy increases if $1<\eta<\infty$ and the Bekenstein-Hawking entropy decreases if $0<\eta<1$.

Case (ii) If
\begin{equation}
\eta=\frac{(1+2a_k+2cmc_uc_r)+\sqrt{(1+2a_k)[(1+2a_k)+2cm c_u(2c_r+ c_u)]}}{[(1+2a_k)-2cm c_r^2]},
\end{equation}
 Eq. (99) represents the change in Bekenstein-Hawking entropy of KNdS given in Eq. (70). When $\eta=1$, Eq. (99) will be the original change in Bekenstein-Hawking entropy $(\eta\Delta S_{BH}=\Delta S_{BH})$ . The change in Bekenstein-Hawking entropy increases $(\eta\Delta S_{BH}>\Delta S_{BH})$ or decreases $(\eta\Delta S_{BH}<\Delta S_{BH})$ respectively if $\eta\in(1,\infty)$ or $\eta\in(0,1)$.

Case (iii) Eq. (99) will be the change in Bekenstein-Hawking entropy given in Eq. (89) if
\begin{equation}
\eta =\frac{[{(1+2a_k)+2cm c_T c_r}+\sqrt{(1+2a_k)[(1+2a_k)+2cm c_T (2c_r-c_T)}]}{
{(1+2a_k)+2cm c_r^2}}.
\end{equation}

If $\eta=1$, the Lorentz violation is cancelled and Eq. (99) becomes $\Delta S_{BH}= (r_f^2 + a^2)/\Xi - (r_i^2 +a^2)/\Xi$, which is the actual change in Bekenstein-Hawking entropy near the event horizon of KNdS black hole.  If $0<\eta<1$ or $1<\eta<\infty$, the change in Bekenstein-Hawking entropy decreases or increases near the event horizon of the KNdS black hole. The variation of change in Bekenstein-Hawking entropies near the event horizon of KNdS black hole are dependent on the choice of ether like vectors $u^{\alpha}$. If $q=0$ in Eq. (99), similar result is obtained in [55]. When $\Lambda=0$ in Eq. (99), our result is consistent with earlier work [56]. If $\eta=1$ and $q=0$ in Eq. (99), our finding is consistent with earlier paper [59]. 

In this paper, we investigate the tunneling of fermions near the event horizon of KNdS black hole in frame dragging coordinate, well-behaved Eddington coordinate system and Painleve coordinate system by using Dirac equation with Lorentz symmetry violation in curved space, Pauli sigma matrices and Feynman prescription.
Under the influence of Lorentz violation theory, the Hawking temperature, heat capacity and change in Bekenstein-Hawking entropy are modified from the actual expressions. In the absence of Lorentz violation theory, the modified Hawking temperatures, heat capacities and change in Bekenstein-Hawking entropies approach to the original Hawking temperature, heat capacity and change in Bekenstein-Hawking entropy near the event horizon of KNdS black hole. The variation of modified Hawking temperatures near the event horizon of KNdS black hole from the original Hawking temperature and the modified heat capacities from the original one have been shown graphically by assigning different values at constant parameters. The existence of phase transition in KNdS black hole is also shown graphically. It is worth mentioning that the position of phase transition does not affect due to Lorentz violation theory. This shows that the modified Hawking temperatures, heat capacities and change in Bekenstein-Hawking entropies depend on choices of ether like vectors $u^{\alpha}$.

\textbf{Acknowledgements}: The first author acknowledges the Council of Scientific and Industrial Research, New Delhi, India, vide letter No. 211610101868 for financial support.

\end{document}